# Habitable Worlds Observatory Living Worlds Science Cases: Research Gaps and Needs

**Co-Authors:** Niki Parenteau, Giada Arney, Eleanora Alei, Ruslan Belikov, Svetlana, Berdyugina, Dawn Cardace, Ligia Coelho, Kevin Fogarty, Kenneth Gordon, Jonathan Grone, Natalie Hinkel, Nancy Kiang, Ravi Kopparapu, Joshua Krissansen-Totton, Emilie LaFleche, Jacob Lustig-Yaeger, Eric Mamajek, Avi Mandell, Taro Matsuo, Connor Metz, Mark Moussa, Stephanie Olson, Lucas Patty, Bill Philpot, Sukrit Ranjan, Edward Schwieterman, Clara Sousa-Silva, Anna Grace Ulses, Sara Walker, Daniel Whitt.

Executive Summary
**The Habitable Worlds Observatory (HWO) is the first astrophysics flagship mission with a key cross-divisional astrobiology science goal of searching for signs of life on rocky planets beyond our solar system.** The Living Worlds Working Group under the Science, Technology, and Architecture Review Team (START) was charged with investigating how HWO could characterize potentially habitable exoplanets orbiting stars in the solar neighborhood, search for signs of life, and interpret potential biosignatures within a false positive and false negative framework. In particular, we focused on (1) identifying biosignatures that have spectral features in the UV-Vis-NIR wavelength range and defining their measurement requirements, (2) determining additional information needed from the planet and planet system to interpret biosignatures and assess the likelihood of false positives, and (3) assembling current knowledge of likely HWO target stars and identify which properties of host stars and systems are most critical to know in advance of HWO. **The Living Worlds atmospheric biosignatures science case is considered one of the key drivers in the design of the observatory**. An additional 10 astrobiology science cases were developed that collectively revealed key research gaps and needs required to fully explore the observatory parameter space and perform science return analyses. **Investment in these research gaps will require coordination across the Science Mission Directorate and fall under the purview of the new Division-spanning astrobiology strategy.**

**Key Recommendations on Needed Research Investments**:
- *Laboratory and Computational Astrochemistry*: line lists, opacities, reaction rates, collisional parameters, and gas phase kinetics to assess the detectability of potential biosignature gases and their chemical network in the atmospheres of other planets.
- *Target Stars and Systems*: stellar elemental abundances to estimate the elemental composition of rocky planets and assess their potential to host life, NUV/FUV fluxes for stars to predict photochemistry and impact on biology.
- *Surface Biosignatures*: build a larger database of laboratory and field pigment and surface reflectance spectra under different conditions, assessment of false positives, interface with Earth remote sensing communities, modeling to assess detectability.
- *Multi-star planet science*: discover and characterize habitable planets around multi-star systems, assess differences from single star systems in planet formation, evolution, properties, and potential to host life.
- *Life on Different Star-Planet Systems:* develop more science cases for plausible photosynthetic and chemosynthetic life on planets with different compositions and formation processes around stellar types other than our Sun.

- *"Pass-through endorsements"*: The Exoplanet Exploration Program maintains a community-driven [Science Gap List](#) that contains updated recommendations for 2025, complementing the Living Worlds gap lists noted below.

**Introduction/Justification**
**See Arney et al. (2025) "The Habitable Worlds Observatory: Telling the Story of Life in the Universe" white paper for a description of the mission science goals.** The Exoplanet Astrobiology community has developed a driving science case based on the chemical and spectral evolution of Earth as a living planet over geologic time. Observed exoplanets will vary significantly in age, including systems far younger and older than our solar system. Correspondingly, a distant observer viewing the Earth billions of years ago would not see tell-tale signs of oxygen or ozone but relatively high concentrations of carbon dioxide and methane from methanogenic microbes, potentially complemented by a biologically maintained hydrocarbon haze layer. The figure below illustrates simulated spectra of "Earth through time," based on reconstructions from evidence from the geologic record. To maximize our probability of finding life elsewhere, HWO should be built to identify Earth-like planets at any stage of their biogeochemical evolution. *While this "Earth through time" science case addressed the charge to 'explore the parameter space in the design of the observatory,' there are other astrobiology science cases that more fully explore 'life as we know it' and 'life as we don't know it.' However, research gaps were identified in the development of these science cases and there are key needs that must be addressed in order to fully determine whether these cases would define new observatory requirements. These gaps and needs are submitted for consideration of cross-Divisional funding opportunities coordinated by the Astrobiology Program.*

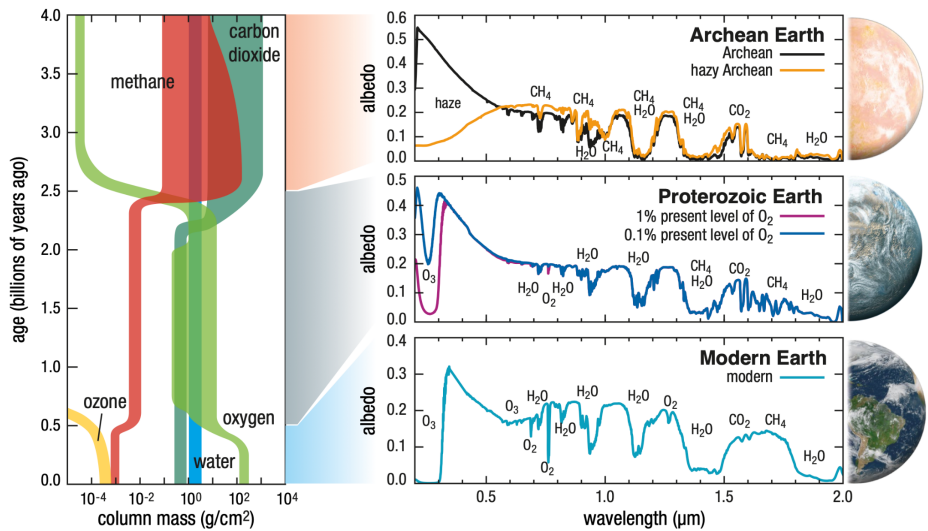

*Figure 1.* Biosignatures and habitability indicators of "Earth through time" reflect how life on our planet has co-evolved with its environment. The biosignatures of Earth through time are a useful minimum set of spectral features to seek on exoplanets with HWO.

*Laboratory and Computational Astrochemistry Gaps/Needs for Biosignatures and Technosignatures*
There is a growing recognition within the exoplanet and astrobiology communities that additional spectroscopic science, both in the laboratory and through computational simulations, is needed to support interpreting observations in the in-term with JWST [**1, 2**] and future missions, including the European ARIEL mission [**3**] and HWO. At some wavelength ranges relevant to HWO, no opacity data exist for some potential atmospheric biosignatures and technosignatures, and without these data, **questions as simple as "how much gas is required to be detectable?" are rendered impossible to answer.** Moreover, kinetics data are essential for assessing the atmospheric lifetimes of many spectrally active trace gases. **Without them, we cannot accurately distinguish between plausible abiotic and biotic sources.** Below, we list specific examples of these science gaps of particular interest to HWO.

*Gaseous Biosignature Gaps/Needs*
- **Potential biosignature opacities:** Several potential biosignature gases lack quantitative opacity measurements at VIS-NIR wavelengths relevant to HWO (~0.4-2.0 μm) including methyl halides (e.g., $CH_3Cl$, $CH_3Br$), polyhalomethanes (e.g., $CHCl_3$, $CHBrCl_2$), organosulfur gases (e.g., $(CH_3)_2S$, $CH_3SH$), and other gases known to be produced by life on Earth including $PH_3$, isoprene ($C_5H_8$), and $(CH_3)Se$. For more abundant potential biosignatures, such as $CH_4$ and $C_2H_2$, adequate line lists exist, but there is very limited data on their collisional broadening parameters for non-Earth-like environments, which will affect altitude and abundance estimates. Finally, very few potential biosignatures have known spectroscopic data (cross-sections and reaction rates) on their formation and destruction network, which severely limits our ability to assess false positive scenarios.
- **Potential technosignature opacities:** Proposed "technosignature" molecules that would fingerprint industrial waste or terraforming processes (climate modification) may have mid-infrared data but lack quality near-infrared and visible opacity data, not just for direct spectra (e.g., line lists, or cross-sections) but also for all other spectroscopic parameters needed for accurate photochemical and radiative transfer modelling. These include fluorine-bearing molecules like chlorofluorocarbons (CFCs; e.g., $CCl_3F$), hydrofluorocarbons (HFCs; e.g., $CHF_3$), and perfluorocarbons (PFCs; e.g., $C_2F_6$), in addition to $NF_3$ and $SF_6$.
- **Kinetics data:** quantifying reaction rates of the gases above with common radicals, including OH, $NO_3$, $O(^1D)$ in bath gases that go beyond Earth-like $N_2$-$O_2$ atmospheres, including $H_2$, He, and $CO_2$ at temperatures 150-400 K.
- **Higher fidelity data for high-priority biosignatures**: We require expanded, high accuracy, broadening parameters and collision-induced absorption (CIA) data for commonly targeted gases, including $O_2$ and $CH_4$. Examples include better parameterized $O_2$-$O_2$ CIA that is critical for distinguishing biotic and abiotic $O_2$ atmosphere scenarios and should be expanded to various plausible collision partners, such as $O_2$-$N_2$ and $O_2$-$CO_2$ at temperatures 150-700K.
- **Infrastructure investment**
  - In general, line lists/opacities will benefit the most from theoretical work but both broadening parameters and reaction rates (useful for networks) will need laboratory investment (theory cannot quite deliver at present).
  - Development of databases not typically supported by ROSES program elements that provide access and community standards for opacity data: Molecules and Atoms in Exoplanet Science: Tools and Resources for Opacities) database (MAESTRO) [**4**].
  - Access to supercomputing resources to compute opacities (ab initio studies).
  - Longer-term sustainable funding to develop a pipeline for training the next generation of researchers to perform theoretical ab initio work and lab spectroscopy work.

*Target Stars and Systems Gaps/Needs*
- Improved characterization of stellar parameters and determination of possible stellar binarity or multiplicity, since those physical parameters influence the habitable zone and dynamical orbital stability zones.
- Stellar elemental or molecular abundance measurements for all targets to 10% precision, specifically for elements:
  - Important to characterize planetary interiors, namely Fe, Mg, Si, which – with O make up ~97% of rocky Solar System planets, as well as the highly refractory Al, Ca, and Ti [**5, 6**].
  - Needed to be available to biology, such as major biological elements: CHNOPS, Na, Mg, Cl, K, Ca, Br; or elements that are essential/beneficial for many organisms: Si, Cr, Mn, Fe, Co, Ni, Cu, Zn, Mo, Cd, W [**7, 8**].
- NUV/FUV fluxes for stars to determine planetary photochemistry and its impact on biology, particularly surface-dwelling photosynthetic microbes, the potential presence and variability of $O_2$/$O_3$ (semi-respectively), and effects on biosignature interpretation.
- Ages of stars to high precision (ideally to < 1 Gyr) to characterize young stars to determine prebiotic chemistry. Understanding stellar ages would impact the interpretation of planet spectra, modeling of exoplanet atmospheres and interiors, placing planets in evolutionary context, searching for signs of oxygenation of habitable worlds as function of age. Stellar properties that help inform age determinations are: stellar rotation periods, X-ray luminosities and X-ray fractional luminosities (log $L_x/L_{bol}$), and stellar abundances of particular elements (Y, Ba, Cs).

*Surface Biosignatures Gaps/Needs*
- Lab measurements of microbial cultures under different conditions: radiation (e.g., UV flares), temperature, pressure broadening, viewing angle.
- Field measurements of reflective surface features of different environments (e.g., natural microbial communities, higher organisms).
- Measurement of homochiral biosignatures in pigment reflectance spectra using full Stokes linear and circular polarization spectropolarimetry.
- False positives: measurements of abiotic surface types, especially ones that may mimic pigment edge features (leveraging other resources like the USGS Spectral Library).
- Integration with Earth observations measuring pigments on continents, near-shore areas, and open ocean blooms, and future observing campaigns (e.g., satellite observations, I.S.S., moonshine, Earthshine, proposed telescopes on the Moon).
- Modeling: (a) 3D GCM studies across multiple planetary parameters (Sun-like stars, K stars, stellar flux, planet size, rotation period, obliquity, eccentricity, atmospheric composition, land/ocean configuration) ported to observation simulation studies (limits of HWO detection and HWO target stars); (b) Time-variability in surface biosignatures and impacts on detections; (c) Self-consistent climate-biota scenarios for detectability, realistic surface spectra; (d) Systematic examination of degeneracies between absorbers (atmospheric and biotic/abiotic surface species) at broad wavelength coverage; (e) Revisit "Earth through time" with more complex models to address the above.

### *Science of Potentially Habitable Planets Around Multi-Star Systems Gaps/Needs*
- HWO targets Sun-like stars, and more than half of these are in multi-star systems. Any census of nearby habitable planets will have a large gap if binary stars are excluded from target lists. In particular, a *complete* census of all nearby potentially habitable planets around Sun-like stars requires us to observe multi-star systems. Furthermore, the nearest possible Sun-like host for potentially habitable planets is the Alpha Centauri system, which contains two Sun-like stars (each with a stable habitable zone).
- Our understanding of potentially habitable planets currently extends mostly to single-star systems. Habitable planet formation may proceed differently in multi-star systems than around single stars, and while we have some theoretical understanding of these differences, there remains a gap both in our theoretical understanding and observational data.
- Habitable planet orbit evolution is affected by the presence of stellar companions. Theoretical models of this are fairly mature, but the observational data is still scarce.
- There is a gap in our knowledge of the amount and type of circumstellar disks in multi-star systems, as well as how they form and interact with planets, compared to single stars. Disks are relevant to habitable planets because they can affect habitability both in positive ways (e.g., water delivery by asteroids and comets), and negative ways (e.g. extinctions of life due to heavy bombardments).

### *Life on Different Star-Planet Systems*
- Develop more science cases for plausible photosynthetic and chemosynthetic life on planets with different formation processes/compositions around other types of stars; assess probability of evolution of different metabolisms on these other systems; leverage other areas of expertise in astrobiology, such as systems biology, to address these topics.

### *"Pass-through endorsements"*
The NASA Exoplanet Exploration Program (ExEP) carries out science and technology research tasks that advance NASA's science goal to "Discover and study planets around other stars, characterize their properties, and identify candidates that could harbor life." The 2025 ExEP Science Gap List section *2.16. SCI-16: Complete the inventory of remotely observable exoplanet biosignatures and their false positives* includes a list of topics that are complementary to the above Living Worlds detailed gap list.